\documentclass[aps,prl,twocolumn,superscriptaddress,a4paper,english,longbibliography,floatfix]{revtex4-2}
\usepackage{times}
\usepackage{url}
\usepackage{soul}
\usepackage{graphicx}
\usepackage{amsmath}
\usepackage{subfigure}
\usepackage{xcolor}
\usepackage{amssymb}
\usepackage{latexsym}
\DeclareGraphicsExtensions{.png,.eps}
\usepackage{makecell}
\usepackage{diagbox}
\usepackage{setspace}
\usepackage{appendix}
\usepackage{etoolbox}

\makeatletter
\newenvironment{suppwidetext}{%
  \par\ignorespaces
  \onecolumngrid
  \vskip10\p@
  \vskip6\p@
  \prep@math@patch
}{%
  \par
  \vskip6\p@
  \vskip8.5\p@
  \twocolumngrid\global\@ignoretrue
  \@endpetrue
}
\makeatother

\newcommand {\R}[1]{#1}

\usepackage[ruled,vlined,linesnumbered]{algorithm2e}
\usepackage[urlcolor=blue]{hyperref}      
\hypersetup{
	colorlinks = true,                    
	citecolor = {blue},
	linkcolor = {purple},
}
\begin{document}	

\title{\R{Diagonalizing large-scale quantum many-body Hamiltonians using variational quantum circuit and tensor network}}

\author{Peng-Fei Zhou}
\affiliation{Center for Quantum Physics and Intelligent Sciences, Department of Physics, Capital Normal University, Beijing 10048, China}
\author{Shuang Qiao}
\affiliation{Center for Quantum Physics and Intelligent Sciences, Department of Physics, Capital Normal University, Beijing 10048, China} 
\author{An-Chun Ji}
\affiliation{Center for Quantum Physics and Intelligent Sciences, Department of Physics, Capital Normal University, Beijing 10048, China} 
\author{Shi-Ju Ran} \email[Corresponding author. Email: ] {sjran@cnu.edu.cn}
\affiliation{Center for Quantum Physics and Intelligent Sciences, Department of Physics, Capital Normal University, Beijing 10048, China}
\date{\today}

	\begin{abstract}
			\R{Exact diagonalization (ED) provides complete access to many-body eigenenergies and eigenstates, yet its exponential cost confines it to small systems. We propose tensor network variational diagonalization (TNVD), which encodes the full eigenenergy spectrum in a matrix product state (MPS) while representing the corresponding eigenstates via a finite depth variational quantum circuit (VQC) acting on product states. TNVD thereby reduces diagonalization complexity from exponential to polynomial in system size \(N\). For the quantum Ising chain, TNVD accurately reproduces eigenenergies for \(N\leq 16\) and, at sizes inaccessible to ED such as \(N=100\), directly samples them from a single MPS encoding all \(2^N\) levels. A random label control shows that this compact \(N\)-site representation depends on how the eigenenergies are organized in label space. TNVD further reveals that, at their integrable limits, the random field Ising and XXZ chains show comparable level spacing ratios and mean eigenstate entanglement entropies but markedly different Schmidt spectrum decay. This difference exposes distinct eigenstate entanglement structures that govern their classical simulability and the difficulty of finite depth quantum circuit preparation. Our work establishes TNVD as a scalable full spectrum diagonalization framework and its VQC as a quantum route to volume-law-entangled eigenstates that challenge efficient classical simulation.}
	\end{abstract}

\maketitle

\begin{figure}[t]
	\centering
	\includegraphics[width=0.38\textwidth]{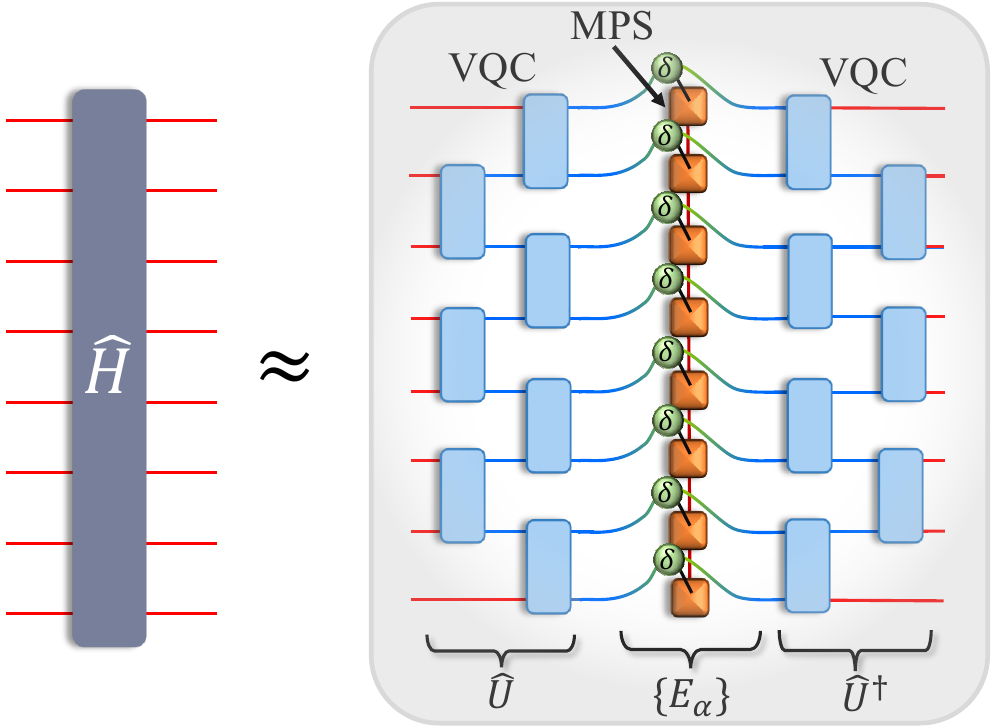}
		\caption{\R{(Color online) TNVD representation of the eigenvalue decomposition in Eq.~(\ref{eq-basis_H}). A spectrum MPS encodes the exponentially many eigenenergies, while a finite depth variational quantum circuit (VQC) maps binary product states to approximate eigenstates. Superdiagonal tensors connect the shared eigenpair labels to the VQC and its conjugate.}}
	\label{fig-Schem}
\end{figure}

\begin{figure*}[t]
	\centering
	\includegraphics[width=1.0\textwidth]{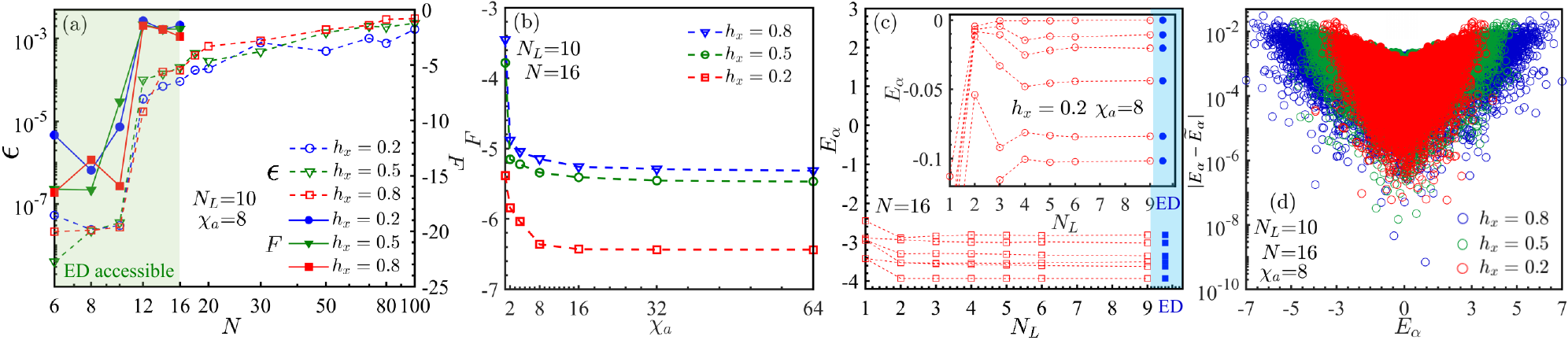}
		\caption{\R{(Color online) Full spectrum simulation for the quantum Ising chain. (a) The mean absolute eigenenergy error $\epsilon$ [left y-axis, Eq.~(\ref{eq-epsilon})] and logarithmic Hilbert--Schmidt loss $F$ [right y-axis, Eq.~(\ref{eq-Schmidt})] versus system size $N$. $\epsilon$ is shown only for $N\leq16$, where ED is available. (b) $F$ versus the spectrum MPS bond dimension $\chi_a$. (c) The lowest few eigenenergies and, in the inset, eigenenergies near $E_\alpha=0$ converge toward ED with circuit depth $N_L$ for $N=16$ and $h_x=0.2$. (d) Level resolved absolute eigenenergy error $|E_\alpha-\tilde E_\alpha|$ versus $E_\alpha$ for $N=16$. Results are shown for $h_x=0.2,0.5,0.8$, except in (c), and unless varied we use $N_L=10$, $\chi_a=8$, and $\chi_t=16$.}}
	\label{fig-fig1}
\end{figure*}


\textit{Introduction.---} \R{The complete eigenstructure of a many-body Hamiltonian is central to quantum many-body physics, determining its thermodynamic properties and dynamical response while providing key signatures of thermalization, quantum chaos, and localization~\cite{ETHDeutschPRA1991,chaosETHPRE1994,statisticalAlessioAP2016}. Exact diagonalization (ED) provides direct access to this eigenstructure across a broad class of many-body systems, but the Hilbert space dimension grows exponentially with system size. Even after exploiting symmetries, interacting spin chains are typically limited to fewer than about thirty spins~\cite{EDLancozsDagottoRMP1994,EDmaxlargeLuitzPRB2015}. The central challenge is therefore to access structure across the full spectrum at sizes beyond ED.}

\R{Tensor network methods, such as matrix product states (MPS), are well established for ground states and have been extended to selected excited eigenstates, particularly in localized regimes~\cite{S11DMRGRev,PollmanRGexcited2016PRL,BryanKexcitedDMRGPRL2017}. In localized systems, quasi-local integrals of motion underpin compact representations of complete eigenbases~\cite{EEMBLSerbynPRL2013,HuseFMBLPRB2014}, while spectral and unitary tensor network constructions encode approximate full eigenbases or diagonalizing transformations~\cite{SpectralTNSChandranPRB2015,SBRGYouPRB2016,Pollmann2016,WahlTNfull2017PRX}. Yet direct access to the complete many-body eigenstructure remains a major challenge at sizes beyond ED.}

\R{In this work, we address the exponential wall in the full diagonalization of many-body Hamiltonians by proposing tensor network variational diagonalization (TNVD). Our key idea is to encode the \(2^N\)-level eigenenergy spectrum in an MPS while representing the corresponding eigenstates with a finite depth variational quantum circuit (VQC), thereby reducing the contraction cost from exponential to polynomial in \(N\). For the quantum Ising chain, TNVD accurately reproduces eigenenergies for \(N\leq16\) and, at \(N=100\), directly samples them from a single spectrum MPS encoding all \(2^N\) levels. A random label control shows that this compact \(N\)-site representation depends on how the eigenenergies are organized in label space. TNVD further reveals that, at their integrable limits, the random field Ising and XXZ chains show comparable level spacing ratios and mean eigenstate entanglement entropies but markedly different Schmidt spectrum decay. This contrast exposes distinct eigenstate entanglement structures that govern their classical simulability and the difficulty of finite depth quantum circuit preparation. Together, these results establish TNVD as a scalable framework for full spectrum diagonalization, with its VQC providing a quantum route to volume-law-entangled eigenstates that challenge efficient classical simulation.}

\textit{Tensor network variational diagonalization approach.---}
Considering a system of $N$ spin-$1/2$ degrees of freedom, the eigenvalue decomposition of its Hamiltonian $\hat{H}$ reads
\begin{equation}\label{eq-basis_H}
	\hat{H} = \sum_{\alpha=0}^{2^N-1} E_\alpha |\alpha\rangle \langle \alpha|,
\end{equation}
where $\{E_\alpha\}$ form the eigenenergy spectrum and $\{|\alpha \rangle\}$ represent the eigenstates satisfying $\hat{H} |\alpha\rangle = E_\alpha |\alpha\rangle$.
The index $\alpha$ can be expressed in the binary form as $\alpha \equiv (r_1, r_2, \cdots, r_N)$ with $r_n = 0$ or $1$ $(n=1,\ldots,N)$.
Then we write $|\alpha\rangle$ as the evolution by the unitary transformation $\hat{U}$ on the product state $\{|r_{\alpha} \rangle \equiv |r_1 r_2 \cdots r_{N}\rangle\}$ as $|\alpha\rangle = \hat{U} |r_{\alpha}\rangle$.

By treating the eigenenergy spectrum as an $N$th-order tensor $E_{r_1r_2 \cdots r_N}$, we encode it into an MPS as
\begin{equation}\label{eq-MPS}
	E_{r_1\cdots r_N}
	=
	\sum_{a_1 \cdots a_N}
	\left(
	\prod_{n=1}^{N}
	A^{[n]}_{r_n a_n a_{n+1}}
	\right),
\end{equation}
with $\{\boldsymbol{A}^{[n]}\}$ the local tensors (see the yellow squares in Fig.~\ref{fig-Schem}) and $\{a_n\}$ the virtual indexes, where $\dim(a_1)=\dim(a_{N+1})=1$ for open boundaries.
The complexity of MPS just scales linearly with $N$ as $O(2N\chi_a^2)$, with the virtual dimension $\chi_a = \dim(a_n)$ a hyperparameter.

\R{The key question is whether the eigenenergy tensor \(E_{r_1\cdots r_N}\) can be efficiently represented as an MPS. A random ordering of the \(2^N\) eigenenergies produces volume law virtual entanglement, requiring the bond dimension \(\chi_a\) to grow exponentially with \(N\)~\cite{PageRandomStatePRL1993,CiracTNRMP2021,RandomPermutation2026}. However, the level ordering is a gauge freedom of the eigendecomposition of Eq.~\eqref{eq-basis_H}, because the eigenenergies and corresponding eigenstates can be permuted together without changing the decomposition. Our calculations show that a level ordering whose virtual entanglement entropy obeys an area law emerges naturally from optimization of the spectrum MPS in Eq.~\eqref{eq-MPS} (see Supplemental Material~\cite{SupplementalMaterial}).}

The unitary transformation $\hat{U}$ is represented as a VQC~\cite{CerezoVQANRP2021,BhartiVQARMP2022}. We use a brick wall circuit of two-body gates (blue squares in Fig.~\ref{fig-Schem}), containing $O(NN_L)$ gates at circuit depth $N_L$. Third-order superdiagonal tensors, $\delta_{abc}=1$ for $a=b=c$ and zero otherwise, connect the spectrum MPS to the VQC and its conjugate~\cite{SchmidtTNSPRL2023}. Together, these tensors form the variational representation of the eigenvalue decomposition of $\hat H$.

The local MPS tensors and VQC gate parameters form the variational parameters of the TNVD ansatz. We use latent tensors~\cite{ZHR21ADQC,PRXRITECQC2021} and obtain the corresponding unitary VQC gates by an SVD-based polar projection. \R{These parameters are optimized by gradient descent to minimize the logarithmic Hilbert--Schmidt loss}
\begin{equation}\label{eq-Schmidt}
    \R{F = \log_2 \left\|\hat{H} - \tilde{H} \right\|_{\mathrm{HS}}^2 - N,}
\end{equation}
\R{where $\tilde{H}$ is the Hamiltonian represented by the ansatz and $\|\cdot\|_{\mathrm{HS}}$ denotes the Hilbert--Schmidt norm~\cite{SchmidtdisVedralPRA1998,SchdistancePRA2019}. Because $\|\hat H-\tilde H\|_{\mathrm{HS}}^2=\mathrm{Tr}[(\hat H-\tilde H)^\dagger(\hat H-\tilde H)]$ measures the global operator residual by tracing over the full many-body Hilbert space, the optimization constrains the Hamiltonian globally rather than privileging a selected energy window.}

To efficiently evaluate $F$, we represent $\hat H$ as a \R{matrix product operator (MPO)} using the \textit{automata} construction~\cite{CroautomataMPOpra2008,Neb_automata_MPO_pra2010}. The three terms in $\|\hat H-\tilde H\|_{\mathrm{HS}}^2$ are evaluated as tensor network contractions. The norm $\mathrm{Tr}(\hat H\hat H^\dagger)$ is obtained by contracting two MPOs. Unitarity of VQC gives $\mathrm{Tr}(\tilde H\tilde H^\dagger)=\sum_\alpha |E_\alpha|^2$, the norm of the spectrum MPS. For the cross term $-2\,\mathrm{Re}\,\mathrm{Tr}(\hat H^\dagger\tilde H)$, the MPO of $\hat H$ is evolved through $\hat U$ and $\hat U^\dagger$ by TEBD~\cite{TEBD_VidalPRL2003,TEBD_VidalPRL2004} and then contracted with the spectrum MPS and the superdiagonal tensors. The TEBD evolution scales as $O(NN_L\chi_t^3)$, where $\chi_t$ is the retained MPO bond dimension (Schmidt rank cutoff). \R{Thus $N_L$, $\chi_a$, and $\chi_t$ respectively control circuit expressiveness, spectrum MPS compression, and Schmidt truncation during MPO evolution. Thus, the computational complexity scales linearly with \(N\), at least for one-dimensional systems.} 


\textit{\R{Full spectrum simulation.---}} \R{We take the quantum Ising chain in a transverse field as an example.} Its Hamiltonian is $\hat{H}=-\sum_{n=1}^{N-1}\hat{S}_n^z\hat{S}_{n+1}^z-h_x\sum_{n=1}^N\hat{S}_n^x$. For $h_x=0.2$, $0.5$, and $0.8$, Fig.~\ref{fig-fig1}(a) shows $F$ for $6\leq N\leq100$. Where ED is available ($N\leq16$), we also compute the mean absolute eigenenergy error
\begin{equation}\label{eq-epsilon}
		\epsilon = \sum_{\alpha=0}^{2^N-1}|E_{\alpha} - \tilde{E}_{\alpha}| / 2^N,
\end{equation}
	with $\{E_{\alpha}\}$ and $\{\tilde{E}_{\alpha}\}$ obtained by sorting the ED and TNVD energy levels in ascending order, respectively. Consistent trends of $F$ and $\epsilon$ are observed, suggesting $F$ as a valid characterization of error. For $ N > 16 $ where $\epsilon$ become inaccessible, $ F $ continues exhibiting sublinear growth, reaching $ \sim O(10^{-1}) $ for $ N \to 100 $, indicating controlled error for large systems. In these simulations, we take the VQC depth $N_L=10$, the virtual dimension of the MPS $\chi_a=8$, and the dimension cutoff in TEBD $\chi_t=16$.

To demonstrate the validity of MPS on representing the exponentially many eigenenergies, Fig.~\ref{fig-fig1}(b) shows $F$ against the virtual dimension $\chi_a$. Convergence is reached for about $\chi_a>16$, suggesting that $\{E_{\alpha}\}$ exhibits high degree of sparsity as a tensor, which can be accurately written into an MPS with moderate $\chi_a$. This is akin to adopting MPS to represent Schmidt coefficients of quantum many-body states~\cite{SchmidtTNSPRL2023}.

Taking $ N = 16 $ and $ h_x = 0.2 $, the main panel of Fig.~\ref{fig-fig1}(c) shows that the low-lying eigenenergies obtained by TNVD converge to those obtained by ED as the VQC depth $ N_L $ increases. Similar observation is made for the eigenenergies at $E_{\alpha} \simeq 0$, as shown in the inset. These imply a VQC with moderate depth can accurately encode the eigenstates. Fig.~\ref{fig-fig1}(d) shows $\R{|E_\alpha-\tilde{E}_\alpha|}$ versus the eigenenergy $E_{\alpha}$ with errors of $O(10^{-2})$ or less, suggesting excellent accuracy. 

\begin{figure}[tbp]
	\centering
		\includegraphics[width=\columnwidth]{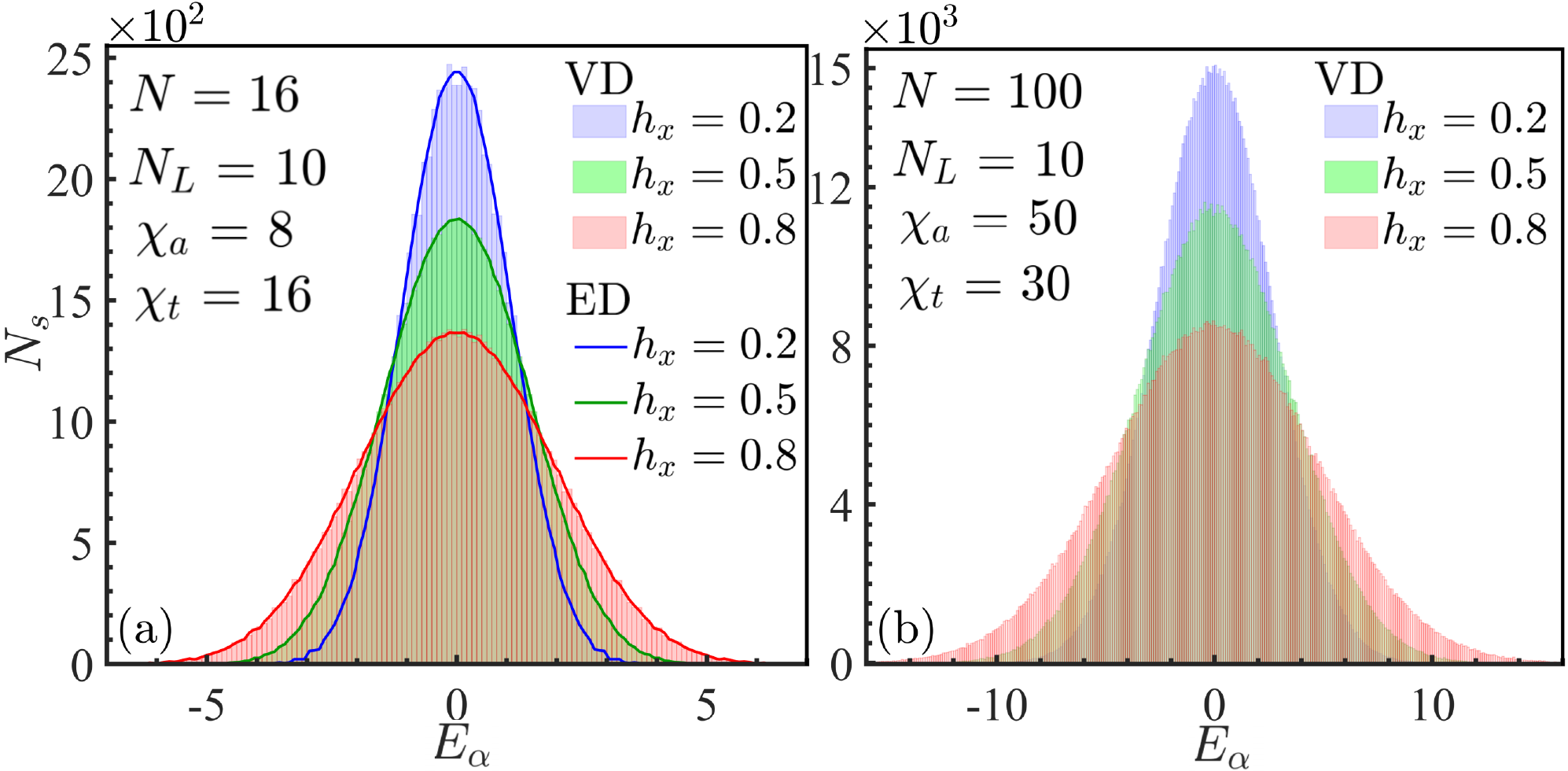}
		\caption{\R{(Color online) Density of states $N_s(E_\alpha)$ of the transverse field Ising chain. (a) For $N=16$ ($N_L=10$, $\chi_a=8$, $\chi_t=16$), the TNVD DOS agrees with ED over the full spectrum. Gaussian fits give $\sigma=1.069,1.426,1.910$ from ED and $\sigma=1.068,1.425,1.908$ from TNVD for $h_x=0.2,0.5,0.8$, respectively. (b) For $N=100$ ($N_L=10$, $\chi_a=50$, $\chi_t=30$), TNVD samples $10^6$ eigenenergies from a single spectrum MPS encoding all $2^{100}$ levels, giving Gaussian widths $\sigma=2.659,3.489,4.676$. We fix $\mu=0$ in all fits.}}
	\label{fig-fig2}
\end{figure}

Figure~\ref{fig-fig2}(a) compares the density of states (DOS) $N_s(E_\alpha)$ from ED and TNVD at $N=16$. Both are well described by the Gaussian
\begin{equation}\label{eq-Gaussian}
	f(E) = \frac{A}{\sigma \sqrt{2\pi}} \exp\left(-\frac{(E - \mu)^2}{2\sigma^2}\right),
\end{equation}
where $A$, $\mu$, and $\sigma$ are the amplitude, mean, and standard deviation, respectively. The fitted widths are $\sigma=1.069, 1.426, 1.910$ (ED) and $\sigma=1.068, 1.425, 1.908$ (TNVD) for $h_x=0.2, 0.5, 0.8$, respectively. We fix $\mu=0$ by Hamiltonian symmetry.

\R{Figure~\ref{fig-fig2}(b) shows the TNVD DOS for $N=100$, far beyond the scope of ED. The sampled distributions are well described by Gaussian fits with $\sigma=2.659, 3.489, 4.676$. We extract $10^6$ eigenenergies efficiently from a single spectrum MPS that encodes all $2^{100}$ levels. The spectrum MPS provides eigenenergy samples at a cost of $O(N\chi_a^2)$ per sample, linear in $N$, thereby enabling access to full spectrum statistics. This sampling capability is a key advantage of TNVD over existing variational schemes~\cite{SchrodiDOSPRB2017,SpectralTNSChandranPRB2015,Pollmann2016,Haghshenas2021uTNC,WahlTNfull2017PRX}.}

\begin{figure*}[t]
    \centering
    \includegraphics[width=0.98\textwidth]{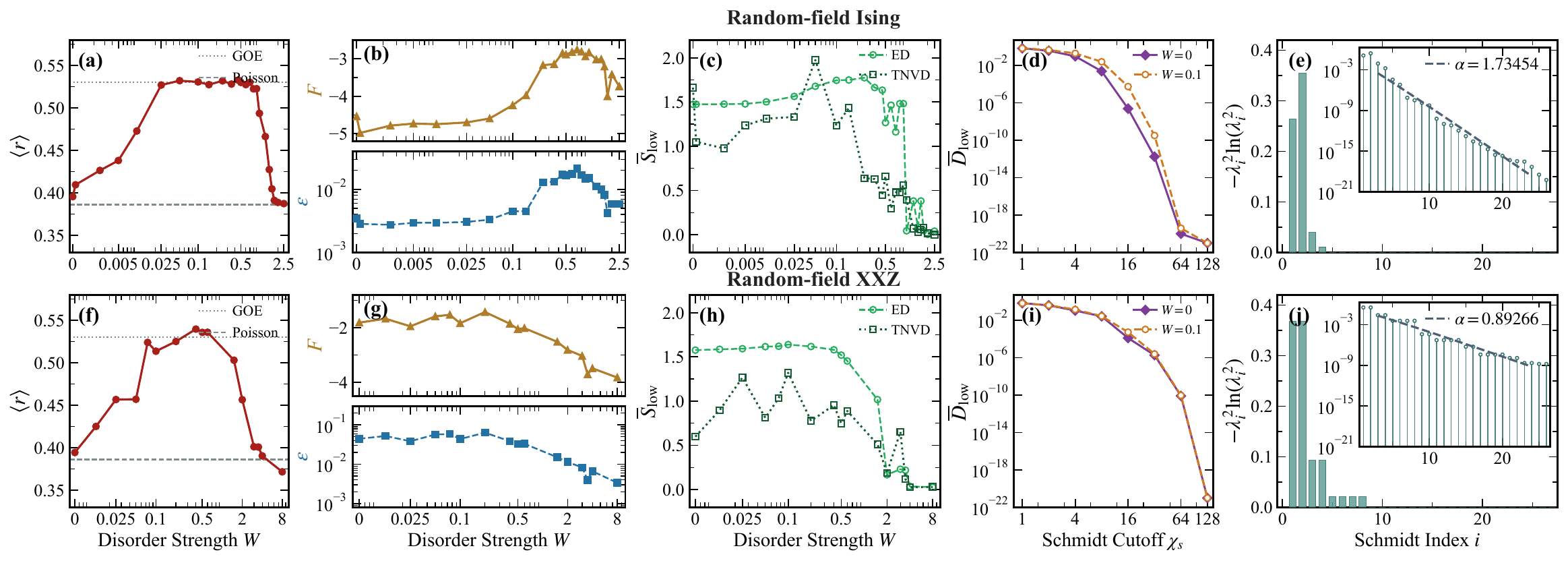}
	   \caption{\R{(Color online) Comparison of the Ising and XXZ chains at $N=14$, with $N_L=10$, $\chi_a=16$, and $\chi_t=48$. The upper and lower rows show the random field Ising ($h_x=0.5$) and isotropic XXZ chains, respectively. (a-c) and (f-h) are shown versus disorder strength $W$. (a,f) Mean adjacent gap ratio $\langle r\rangle$. For XXZ, $\langle r\rangle$ is evaluated in the zero magnetization sector ($k=N/2=7$ up spins), using the central half of that sector. This restriction applies only to $\langle r\rangle$. (b,g) Logarithmic Hilbert--Schmidt loss $F$ and mean eigenenergy error $\epsilon$. (c,h) ED and TNVD mean middle-cut eigenstate entanglement entropy $\overline{S}_{\mathrm{low}}$ over the first 100 excited levels above the ground state in the energy-sorted complete spectrum. (d,i) Average discarded Schmidt weight $\overline{D}_{\rm low}$ versus retained Schmidt rank $\chi_s$, computed from exact ED eigenstates in the same 100-state low-energy window at $W=0$ and $W=0.1$. (e,j) Ground-state entropy contributions $s_i=-\lambda_i^2\ln\lambda_i^2$ at $W=0$, where $\lambda_i$ are the Schmidt coefficients. Insets show the Schmidt weights $\lambda_i^2$ and fitted tails.}}
    \label{fig-biSwithW}
\end{figure*}

\textit{\R{Efficiency and eigenstate entanglement.---}} One key factor that affects the efficiency of TNVD lies in the encoding of eigenstates by VQC. Our main aim below is \R{therefore to uncover the relation between TNVD efficiency and the entanglement structure of eigenstates} by considering \R{the quantum Ising chain above and, for comparison, the isotropic XXZ chain, both subject to random longitudinal fields. The latter is described by $\hat H_{\rm R}^{\rm XXZ}=\sum_{n=1}^{N-1}\sum_{\gamma=x,y,z}\hat S_n^\gamma\hat S_{n+1}^\gamma+\sum_n w_n\hat S_n^z$, where $w_n\in[-W,W]$. At $W=0$, the Ising and XXZ chains are free-fermion and Bethe integrable, respectively, while the random longitudinal fields generically break integrability for $W>0$~\cite{Bethe1931,KarthikTiltedIsingPRA2007,ZnidProsenXXZPRB2008}.} \R{We characterize the disorder-driven level statistics crossover using the mean adjacent gap ratio $\langle r\rangle=\frac{1}{N_{\rm gap}}\sum_n\frac{\min(s_n,s_{n-1})}{\max(s_n,s_{n-1})}$, with $s_n=E_{n+1}-E_n$, where $N_{\rm gap}$ is the number of included spacing ratios. $\langle r\rangle\simeq0.5307$ for Gaussian orthogonal ensemble (GOE) statistics and $\simeq0.3863$ for Poisson statistics~\cite{WignerRandomMatrices1955,DysonThreefoldWay1962,BerryTabor1977,OganesyanMBLPRB2007,DisRatioAtasPRL2013}.} \R{For XXZ, $\langle r\rangle$ is evaluated in the conserved zero magnetization sector, with $k=N/2=7$ up spins for $N=14$, using the central half of that sector's spectrum.}

\R{Figures~\ref{fig-biSwithW}(a,f) show closely parallel level statistics crossovers in the two models, from the integrable Poisson values through a GOE-like regime and back toward Poisson statistics at strong disorder. For both models, we use the same $N_L=10$, $\chi_a=16$, and $\chi_t=48$. Panels (b,g), however, show clearly different TNVD accuracies, with larger $F$ and $\epsilon$ for the XXZ chain. We therefore ask whether this difference is related to their eigenstate entanglement. Panels (c,h) show the ED and TNVD mean eigenstate entanglement entropies over the same 100-state low-energy window. Since the same qualitative contrast persists near the integrable point, we focus on $W=0$ as a representative case, where the two models have comparable mean entanglement entropies despite their different TNVD accuracies. The Supplemental Material further shows how the ED eigenstate entanglement entropy is distributed across the full Ising spectrum as $W$ varies~\cite{SupplementalMaterial}. This raises the question of whether the two models differ in how their entanglement is distributed over the Schmidt spectrum.}

\R{To answer this question, Figs.~\ref{fig-biSwithW}(d,i) show the average discarded Schmidt weight $\overline{D}_{\rm low}$ computed from exact ED eigenstates in the same 100-state low-energy window at $W=0$. It measures the Schmidt weight beyond the retained rank $\chi_s$, averaged over these eigenstates~\cite{TEBD_VidalPRL2003,schollwock2011,EisertAreaRMP2010,S11DMRGRev,SupplementalMaterial}. The decay of $\overline{D}_{\rm low}$ with $\chi_s$ therefore directly probes the Schmidt spectrum tail. The discarded weight decays much more slowly in the XXZ chain. At $\chi_s=8$, it is about one order of magnitude larger than in the Ising chain, and at $\chi_s=16$ the difference reaches several orders of magnitude. Panels (e,j) provide a further comparison through the ground-state Schmidt spectrum. Fitting $\lambda_i^2=C e^{-\alpha i}$ over $3\leq i\leq24$ gives $\alpha\simeq0.893$ for XXZ and $\alpha\simeq1.735$ for Ising, showing that the XXZ Schmidt spectrum has a much heavier tail. Thus, although the two models have comparable mean entanglement entropies at $W=0$, their entanglement is distributed very differently across the Schmidt spectrum.}

\R{The entropy and Schmidt spectrum results therefore reveal that the Ising and XXZ chains possess distinctly different eigenstate entanglement structures, even when their mean entanglement entropies are comparable. This distinction is consistently reflected in both the low-energy discarded weights and the ground-state Schmidt tails. At $W=0$, the isotropic XXZ and critical Ising chains belong to different conformal field theories, with central charges $c=1$ and $c=1/2$, respectively. Since the central charge determines the universal distribution of Schmidt weights at criticality, this difference provides a direct physical basis for the distinct Schmidt spectrum structures observed here, including the heavier XXZ tail~\cite{CalabreseLefevrePRA2008,TagliacozzoFiniteChiPRB2008,PollmannFiniteEntPRL2009,PollmannMooreNJP2010}. These different eigenstate entanglement structures are reflected in the contrasting TNVD efficiencies and are directly relevant to both classical tensor network simulability and finite depth quantum circuit preparation.}

\textit{Summary.---} \R{We have introduced TNVD, a scalable framework for full spectrum diagonalization that represents the full many-body eigenenergy spectrum by an MPS and the corresponding eigenstates by a finite depth VQC. It accurately reproduces the ED eigenenergies for $N\leq16$ and, for sizes beyond ED such as $N=100$, directly samples from a single MPS encoding all $2^{100}$ energy levels. The Ising--XXZ comparison further reveals that, at their integrable limits, the two models have comparable level statistics and mean eigenstate entanglement but markedly different Schmidt spectrum decay. This contrast exposes distinct eigenstate entanglement structures that are reflected in their different TNVD efficiencies and connect full spectrum diagonalization to the classical simulability of many-body eigenstates. The VQC component of TNVD further provides a natural route toward eigenstates whose entanglement structure challenges efficient classical tensor network representations.}

\R{The source code implementing TNVD is publicly available in Ref.~\cite{TNVDGitHub2026}.}

\textit{Acknowledgment.}
\R{PFZ thanks Guo-Dong Cheng, Hao Zhu, and Ding-Zu Wang for stimulating discussions.} \R{PFZ acknowledges support from the Doctoral Student Special Program of the Young Elite Scientists Sponsorship Program by the China Association for Science and Technology (CAST).} This work was supported in part by the NSFC (Grant No. 62175169 and No. 12404092) and Beijing Natural Science Foundation (Grant No. 1232025). \R{SJR acknowledges support from the Peng Huanwu Visiting Professor Program, Chinese Academy of Sciences, and the Academy for Multidisciplinary Studies, Capital Normal University.} The numerical simulations were partially performed on the robotic AI-Scientist platform of Chinese Academy of Sciences.

\clearpage
\setcounter{section}{0}
\setcounter{figure}{0}
\setcounter{equation}{0}
\setcounter{algocf}{0}
\renewcommand{\thesection}{S\arabic{section}}
\renewcommand{\thefigure}{S\arabic{figure}}
\renewcommand{\theequation}{S\arabic{equation}}
\renewcommand{\thealgocf}{S\arabic{algocf}}
\renewcommand{\theHsection}{supp.\arabic{section}}
\renewcommand{\theHfigure}{supp.\arabic{figure}}
\renewcommand{\theHequation}{supp.\arabic{equation}}
\newcommand{\suppsection}[1]{%
  \refstepcounter{section}%
  \section*{\thesection.\quad #1}%
  \addcontentsline{toc}{section}{\thesection.\quad #1}%
}

\begin{suppwidetext}
\begin{center}
{\large\bfseries Supplemental Material for \textquotedblleft{}Diagonalizing large-scale quantum many-body Hamiltonians using variational quantum circuit and tensor network\textquotedblright{}}
\end{center}
\begin{center}
\begin{minipage}{0.92\textwidth}
\small
\textbf{Abstract.}
\R{This Supplemental Material provides additional analyses and numerical details for tensor network variational diagonalization (TNVD). We analyze the virtual entanglement of the eigenenergy tensor in binary label space and present the random label control used to examine the condition for an efficient spectrum MPS representation. We then give the layerwise TNVD optimization algorithm and the disorder protocol. For the comparison between the Ising and XXZ chains in the Letter, we define the low-energy mean eigenstate entanglement entropy and average discarded Schmidt weight, clarify the distinct roles of $\chi_s$ and $\chi_t$, and examine the convergence of the TEBD evaluation with $\chi_t$. Finally, we present full spectrum ED eigenstate entanglement maps for the random field Ising chain, which complement the low-energy analysis in the Letter.}
\end{minipage}
\end{center}
\end{suppwidetext}

\suppsection{Virtual entanglement of the eigenenergy spectrum}

\R{The Letter identifies how the eigenenergies are organized in label space as a key condition for efficiently representing the eigenenergy tensor by a spectrum MPS, since a random ordering of the $2^N$ eigenenergies produces volume law virtual entanglement and therefore requires the spectrum MPS bond dimension $\chi_a$ to grow exponentially with $N$ at fixed accuracy~\cite{PageRandomStatePRL1993,EisertAreaRMP2010,CiracTNRMP2021}. To test this structure directly, we regard the eigenenergy tensor as a normalized spectrum state}

\begin{equation}
|E\rangle =
\frac{1}{\mathcal{N}}
\sum_{r_1,\cdots,r_N}
E_{r_1\cdots r_N}
|r_1\cdots r_N\rangle ,
\end{equation}

\R{where $\mathcal{N}=(\sum_{r_1\cdots r_N}|E_{r_1\cdots r_N}|^2)^{1/2}$. Across the cut after label site $\ell$, we compute
$S_\ell=-\mathrm{Tr}(\rho_\ell\ln\rho_\ell)$, where $\rho_\ell$ is the reduced density matrix of the first $\ell$ binary indices. This quantity is the virtual entanglement entropy in binary label space rather than the eigenstate entanglement entropy, and it quantifies the entanglement relevant to an efficient spectrum MPS representation across each binary label cut.}

\begin{figure}[!t]

\centering

\includegraphics[width=0.98\columnwidth]{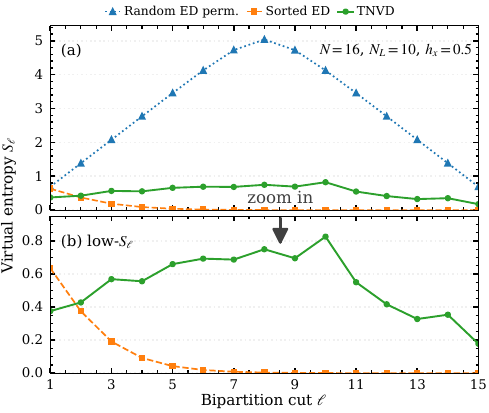}

\caption{\R{Virtual entanglement entropy $S_\ell$ of the normalized spectrum state for the transverse-field Ising chain with $N=16$, $N_L=10$, and $h_x=0.5$. (a) Comparison of the randomly relabeled ED spectrum, the energy-sorted ED spectrum, and the TNVD spectrum with the level ordering obtained by TNVD. Random relabeling changes only the assignment of the exact eigenenergies to binary labels and produces volume law virtual entanglement, whereas the energy-sorted ED and TNVD spectra remain on the low entanglement scale. (b) Zoom of the low-$S_\ell$ regime, resolving the energy-sorted ED and TNVD curves. The random label control lies outside the displayed range. No energy sorting is imposed on the TNVD spectrum.}}

\label{fig-spectrum-random-perm}

\end{figure}

\R{Figure~\ref{fig-spectrum-random-perm} compares three spectrum states for the transverse-field Ising chain at $N=16$, $N_L=10$, and $h_x=0.5$. The randomly relabeled ED spectrum preserves all exact eigenenergy values but changes their assignment to binary labels, producing volume law virtual entanglement with a large middle-cut entropy. The energy-sorted ED spectrum provides an exact reference with low virtual entanglement. By contrast, the TNVD spectrum is shown with the level ordering obtained by TNVD, with no energy sorting imposed, and remains on the low entanglement scale. Panel (b) magnifies this regime and resolves the energy-sorted ED and TNVD curves. Random relabeling leaves the eigenenergy spectrum unchanged while strongly increasing $S_\ell$, directly isolating the role of how the eigenenergies are organized in label space in an efficient spectrum MPS representation. We verified the same qualitative separation at the other accessible sizes examined and for $h_x=0.2$ and $h_x=0.8$. These cases are omitted for compactness.}

\R{Figure~\ref{fig-spectrum-random-perm} therefore supports the statement in the Letter that an efficient spectrum MPS representation depends on how the eigenenergies are organized in label space. The low virtual entanglement found by TNVD, together with the same qualitative behavior at the accessible sizes examined, is consistent with a level ordering whose virtual entanglement entropy obeys an area law, as stated in the Letter. This section addresses the efficient MPS representation of the eigenenergy tensor, while Schmidt truncation during MPO evolution is analyzed separately below.}

\suppsection{Numerical optimization and disorder protocol}

\R{Algorithm~\ref{alg:tnvd} summarizes the layerwise optimization implemented in the public code. We denote the spectrum MPS tensors by \(\Theta_E=\{A^{[n]}\}\), the latent VQC tensors by \(\Theta_U=\{X_{p,j}\}\), and the binary label by \(\mathbf r=(r_1,\ldots,r_N)\), where \(p\) and \(j\) index the circuit layer and two-site gate position, respectively. The optimization stage index \(L\) is zero-based. \(L=0\) is the first optimization stage, and \(\hat U_L\) contains \(L+1\) circuit layers. The update schedule \(\mathcal S\), stopping rule \(\mathcal C\), and optimizer settings are configurable. The corresponding settings are provided with the public code~\cite{TNVDGitHub2026}.}

{\color{black}

\begin{algorithm*}[t]

\small

\DontPrintSemicolon

\caption{Layerwise TNVD optimization}

\label{alg:tnvd}

\KwIn{Hamiltonian MPO \(\hat H\), \(N_L^{\max}\), \(\chi_a\), \(\chi_t\), epoch limit \(T\), update schedule \(\mathcal S\), stopping rule \(\mathcal C\)}

\KwOut{Spectrum MPS \(\Theta_E^\star\) and approximate diagonalizing VQC \(\hat U^\star\)}

\(c_H\leftarrow\operatorname{Tr}(\hat H\hat H^\dagger)\)\;

Initialize the spectrum MPS \(\Theta_E\) with bond dimension \(\chi_a\) and the initial latent VQC tensors \(\Theta_U\)\;

\For{\(L\leftarrow0\) \KwTo \(N_L^{\max}-1\)}{

  Append a gate layer initialized near the identity if \(L>0\), using the tensors optimized at the preceding stage as the initial values\;

  Initialize AdamW and its layer-dependent learning rate schedule\;

  \For{\(t\leftarrow1\) \KwTo \(T\)}{

    Select the active parameter set from \(\{\Theta_E,\Theta_U\}\) according to \(\mathcal S\)\;

    Obtain each unitary VQC gate from its latent tensor by the SVD-based polar projection \(X_{p,j}=Q\Sigma V^\dagger\mapsto u_{p,j}=QV^\dagger\)\;

    Use \(\Theta_E\) as the spectrum MPS representation of \(E_{\mathbf r}\) and set

    \(\hat H_U\leftarrow\operatorname{TEBD}_{\chi_t}(\hat U_L^\dagger\hat H\hat U_L)\)\;

    Contract the tensor networks for

    \(C\leftarrow\sum_{\mathbf r}E_{\mathbf r}(\hat H_U)_{\mathbf r\mathbf r}\)\;

    \(F\leftarrow\log_2[c_H+\sum_{\mathbf r}|E_{\mathbf r}|^2-2\operatorname{Re}C]-N\)\;

    Differentiate \(F\), clip the gradient norm to one, and update the active parameter set\;

    Advance the cosine learning rate schedule\;

    \If{\(\mathcal C(F_{1:t})\) is satisfied}{\textbf{break}\;}

  }

  Save the optimized current stage tensors for the next stage\;

}

\end{algorithm*}

}

\begin{samepage}

\R{All sums in Algorithm~\ref{alg:tnvd} are evaluated by tensor network contractions. The method does not explicitly construct the complete \(2^N\)-level spectrum during optimization.}

\R{For each value of $W$, the random longitudinal fields $w_n$ are independently drawn from a uniform distribution over $[-W,W]$. A separate disorder realization is generated for each $W$. The random seed for each realization is recorded in the code so that the same set of $w_n$ can be generated again. We use one disorder realization at each $W$ and do not average over different realizations.}

\end{samepage}

\suppsection{Schmidt tail analysis for Fig.~4}

\R{The TNVD calculations in Fig.~4 use $N_L=10$, spectrum MPS bond dimension $\chi_a=16$, and retained MPO bond dimension $\chi_t=48$ in the TEBD evolution. Let $\mathcal{L}_{100}(W)$ denote the first 100 excited levels above the ground state in the energy-sorted complete many-body spectrum at disorder strength $W$. The same 100-state low-energy window is used for ED and TNVD. The zero magnetization sector restriction used for the XXZ mean adjacent gap ratio $\langle r\rangle$ applies only to $\langle r\rangle$ and is not used here. For $X\in\{\mathrm{ED},\mathrm{TNVD}\}$, the mean middle-cut eigenstate entanglement entropy shown in Fig.~4(c,h) is defined as
\[
\overline{S}_{\mathrm{low}}^{X}(W)
=
\frac{1}{100}
\sum_{\alpha\in\mathcal{L}_{100}(W)}
S_X^{(\alpha)} .
\]}

\R{For Fig.~4(d,i), we use exact ED eigenstates from the same 100-state low-energy window at $W=0$. Let $\lambda_\mu^{(\alpha)}$ denote the middle-cut Schmidt coefficients of exact eigenstate $\alpha$, ordered such that
$[\lambda_1^{(\alpha)}]^2\geq[\lambda_2^{(\alpha)}]^2\geq\cdots$.
The average discarded Schmidt weight is defined as}

\begin{equation}
\overline{D}_{\mathrm{low}}(\chi_s)
=
\frac{1}{100}
\sum_{\alpha\in\mathcal{L}_{100}(0)}
\sum_{\mu>\chi_s}
\left[\lambda_{\mu}^{(\alpha)}\right]^2 ,
\end{equation}

\R{where $\chi_s$ is the retained Schmidt rank. The quantity $\chi_s$ is used only in the analysis of the exact ED eigenstates and is not a TNVD parameter. It should not be confused with $\chi_t$, which is the retained MPO bond dimension used during TEBD evolution. Therefore, $\overline{D}_{\mathrm{low}}$ characterizes the Schmidt tail structure of the exact eigenstates rather than the Schmidt truncation of the MPO during TEBD evolution.}

\R{Figure~4(e,j) provides a separate ground-state analysis at $W=0$ and is not included in the 100-state average above. The ground-state Schmidt weights are fitted to
$\lambda_i^2=C e^{-\alpha i}$ over $3\leq i\leq24$.}

\suppsection{Convergence of the TEBD evaluation with \texorpdfstring{$\chi_t$}{chi-t}}

\begin{figure}[tbp]

\centering

\includegraphics[width=0.90\columnwidth]{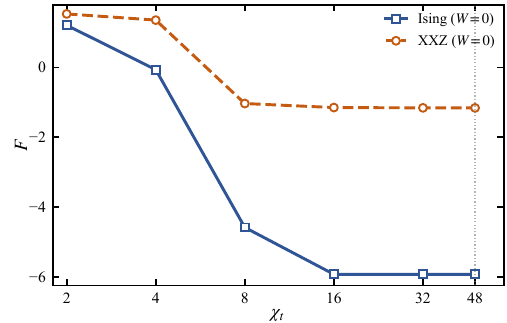}

\caption{\R{Convergence of the logarithmic Hilbert--Schmidt loss $F$ with the retained MPO bond dimension $\chi_t$ for the Ising and XXZ benchmarks at $W=0$, $N=14$, $N_L=10$, and $\chi_a=16$. The optimized TNVD tensors are kept fixed while only $\chi_t$ is varied in the TEBD evaluation. The vertical dotted line marks $\chi_t=48$, which is used in Fig.~4.}}

\label{fig-chit-convergence}

\end{figure}

\R{To test the convergence of the TEBD evaluation of $F$, we keep the optimized TNVD tensors fixed and vary only the retained MPO bond dimension $\chi_t$ in the TEBD evaluation. Figure~\ref{fig-chit-convergence} shows the results for the two $W=0$, $N=14$ benchmarks in Fig.~4, with $N_L=10$ and $\chi_a=16$. For Ising, $F=-5.9317$ at $\chi_t=48$ and differs from its value at $\chi_t=32$ by only $5.2\times10^{-6}$. For XXZ, $F=-1.1568$ at $\chi_t=48$ and the corresponding difference is $7.6\times10^{-4}$. These results show that the reported values of $F$ are numerically stable at $\chi_t=48$. Because the optimized tensors are kept fixed, this test checks only the convergence of the TEBD evaluation and does not test whether reoptimization at larger $\chi_t$ would further improve $F$.}

\suppsection{Full spectrum ED entanglement maps}

\R{To complement the low-energy mean eigenstate entanglement entropy in Fig.~4(c,h), we examine the eigenstate entanglement over the complete spectrum of the $N=14$ random field Ising chain using ED. For an exact eigenstate $|\alpha\rangle$, the middle-cut eigenstate entanglement entropy is defined as}

\begin{equation}
S_\alpha
=
-\operatorname{Tr}
\left[
\rho_A^{(\alpha)}
\ln\rho_A^{(\alpha)}
\right],
\label{eq:full-entropy}
\end{equation}

\R{where $\rho_A^{(\alpha)}$ is the reduced density matrix of one half of the chain. We define the normalized energy density as}

\begin{equation}
\bar E_\alpha
=
\frac{E_\alpha-E_{\min}}
{E_{\max}-E_{\min}}.
\label{eq:normalized-energy}
\end{equation}

\begin{figure}[t]

\centering

\includegraphics[width=0.98\columnwidth]{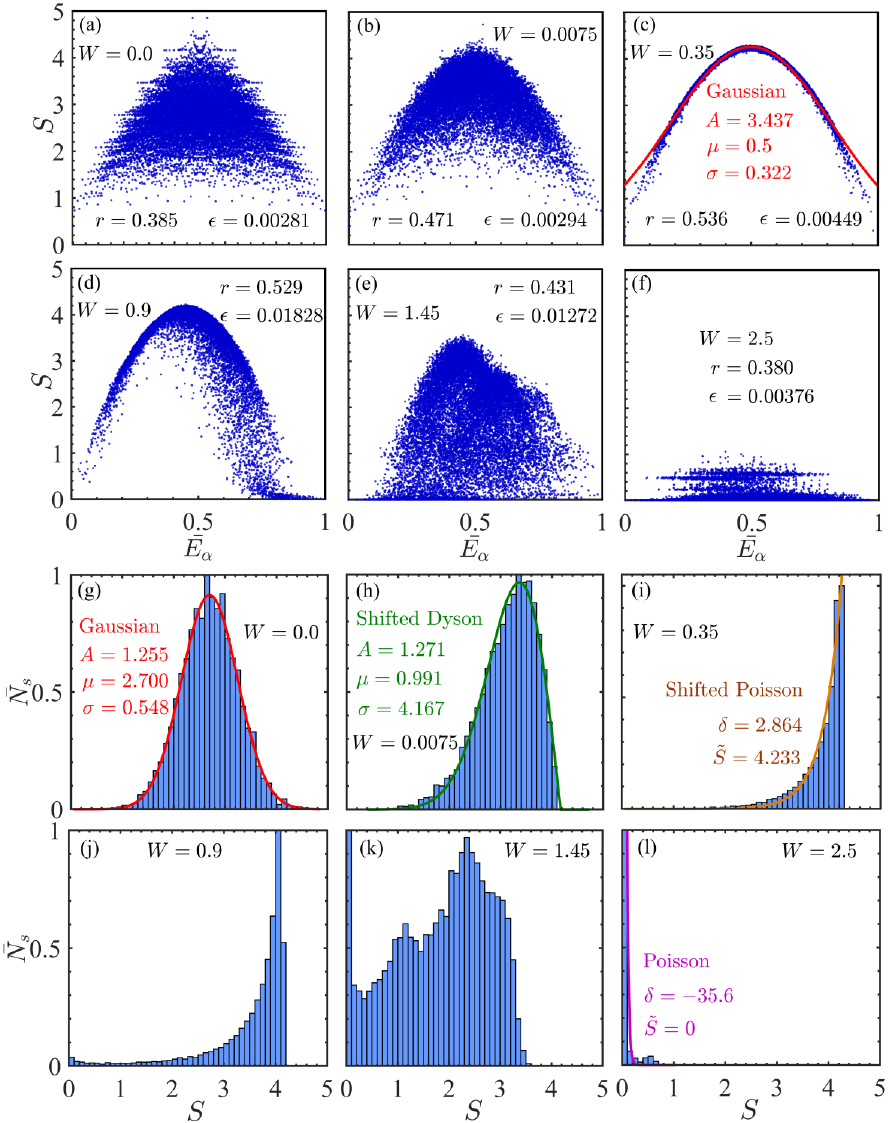}

\caption{\R{Full spectrum ED eigenstate entanglement for the $N=14$ random field Ising chain at the middle cut. Panels (a--f) show $S_\alpha$ versus normalized energy density $\bar E_\alpha$ for $W=0$, $0.0075$, $0.35$, $0.9$, $1.45$, and $2.5$, respectively. Panels (g--l) show the corresponding normalized distributions $\bar N_s(S)$. The fitted curves in panels (c), (g), (h), (i), and (l) use the forms defined in Eqs.~\eqref{eq:full-gaussian}--\eqref{eq:full-poisson}. In panels (a--f), $r$ denotes the mean adjacent gap ratio $\langle r\rangle$ and $\epsilon$ is the mean eigenenergy error defined in the Letter. With increasing disorder, the eigenstate entanglement develops a pronounced arc-like structure across the spectrum and eventually collapses toward low entanglement at strong disorder.}}

\label{fig-full-spectrum-ed}

\end{figure}

\R{For each value of $W$, let $N_s(S)$ be the number of eigenstates in the entropy bin centered at $S$. The normalized distribution shown in Fig.~\ref{fig-full-spectrum-ed} is defined as}

\begin{equation}
\bar N_s(S)
=
\frac{N_s(S)}
{\max_{S'}N_s(S')}.
\label{eq:normalized-state-count}
\end{equation}

\R{Thus, the maximum of $\bar N_s(S)$ is one. The quantity $\bar N_s(S)$ describes how the eigenstates are distributed over entanglement entropy and is distinct from the energy DOS $N_s(E_\alpha)$ in the Letter.}

\R{For $W>0$, the random longitudinal fields generically break the free-fermion integrability of the $W=0$ Ising chain~\cite{KarthikTiltedIsingPRA2007,BanulsThermalizationPRL2011}. The distribution of eigenstate entanglement across the many-body spectrum has been widely studied in disordered interacting systems~\cite{EDmaxlargeLuitzPRB2015,strongWEEBardarsonPRL2014}. All entanglement data in Fig.~\ref{fig-full-spectrum-ed} are obtained from exact ED eigenstates. The quantities $\langle r\rangle$ and $\epsilon$ are included only for comparison with the results in the Letter.}

\R{The curves in panels (c) and (g) are fitted independently by the Gaussian~form}

\begin{equation}
g(x)
=
\frac{A}{\sigma\sqrt{2\pi}}
\exp\left[
-\frac{(x-\mu)^2}{2\sigma^2}
\right],
\label{eq:full-gaussian}
\end{equation}

\R{where $A$ is the amplitude, $\mu$ is the center, and $\sigma$ is the width. In panel (c), $x=\bar E_\alpha$ and the Gaussian describes the dependence of $S_\alpha$ on energy density. In panel (g), $x=S$ and the Gaussian describes the normalized distribution $\bar N_s(S)$. The fit parameters are given directly in the corresponding panels. The pronounced arc in panel (c) is centered near the middle of the spectrum. This arc-like dependence of eigenstate entanglement on energy has also been reported in many-body spectra~\cite{arcIversenPRB2023,arcQianqianPRB2024}.}

\R{The distributions in panels (h), (i), and (l) are fitted by one-sided fitting forms. For panel (h), we use the shifted Dyson~form}

\begin{equation}
\bar N_s(S)
=
A\frac{\pi}{2}\mu(\sigma-S)
\exp\left[
-\frac{\pi}{4}\mu(\sigma-S)^2
\right],
\qquad S<\sigma ,
\label{eq:full-dyson}
\end{equation}

\R{where $A$ is the amplitude, $\sigma$ gives the shift, and $\mu$ controls the width. For panels (i) and (l), we use the exponential form}

\begin{equation}
\bar N_s(S)
=
\exp\left[
-\delta(\widetilde S-S)
\right].
\label{eq:full-poisson}
\end{equation}

\R{Here $\widetilde S$ gives the shift, while the sign and magnitude of $\delta$ determine the direction and slope of the exponential dependence. This form is labeled ``shifted Poisson'' in panel (i) and ``Poisson'' in panel (l). The fit parameters are given directly in the corresponding panels.}

\R{These fits summarize how the full spectrum eigenstate entanglement distributions change with disorder. At intermediate disorder, $S_\alpha$ develops a pronounced arc-like dependence on energy density, with the largest entanglement near the middle of the spectrum. At strong disorder, the distribution collapses toward low $S_\alpha$, consistent with the suppression of eigenstate entanglement in the localized regime~\cite{collapsesdisBayganPRB2015,collapsesLimPRB2016,LongtailcollapsesDavidPRB2016}. The names ``shifted Dyson'' and ``shifted Poisson'' refer only to the fitting forms of $\bar N_s(S)$ and do not assign level statistics to the entropy variable $S$. These full spectrum ED results complement the low-energy averages in Fig.~4 by showing how eigenstate entanglement is distributed across the complete spectrum. They are distinct from the virtual entanglement of the eigenenergy tensor in Fig.~\ref{fig-spectrum-random-perm}.}

\nocite{LongtailcollapsesDavidPRB2016,MBLPhaseGeraedtsNJP2017,MBLPhaseKhemaniPRX2017,strongWEEBardarsonPRL2014}
\bibliography{bibi}
\end{document}